\documentclass[aps,12pt,showkeys,amsmath,amssymb,amsfonts]{revtex4-2}

\usepackage{mathrsfs}%
\usepackage[title]{appendix}
\usepackage{amsthm}
\usepackage{bm}

\def\I{\mathbb{1}}
\def\O{\mathbb{0}}
\renewcommand{\o}{\mathbf{0}}
\newcommand{\be}{\begin{equation}}
\newcommand{\ee}{\end{equation}}
\newcommand{\ba}{\begin{eqnarray}}
\newcommand{\ea}{\end{eqnarray}}
\newcommand{\nn}{\nonumber}

\renewcommand{\d}{\partial}

\newcommand{\q}{\mathbf{x}}
\newcommand{\p}{\mathbf{p}}
\newcommand{\x}{\mathbf{x}}
\newcommand{\y}{\mathbf{y}}

\newcommand{\ca}{\mathbf{a}}

\newcommand{\A}{\bm{A}}
\newcommand{\E}{\bm{E}}
\newcommand{\B}{\bm{B}}

\newcommand{\pib}{\bm\pi}

\renewcommand{\L}{{\parallel}}
\newcommand{\T}{{\perp}}

\begin{document}

\title{Alternative Hamiltonian and Lagrangian Approaches for Constrained Systems}

\author{M.A. de Andrade}\thanks{e-mail: marco@fat.uerj.br}
\author{C. Neves}\thanks{e-mail: clifford@fat.uerj.br}
\author{E.V. Corr\^{e}a Silva}\thanks{e-mail: eduardo.vasquez@fat.uerj.br}
\affiliation{Departamento de Matem\'{a}tica, F\'\i sica e Computa\c{c}\~{a}o, Faculdade de Tecnologia, \\ 
Universidade do Estado do Rio de Janeiro,\\
Rodovia Presidente Dutra, Km 298, P\'{o}lo
Industrial, \\CEP 27537-000, Resende-RJ, Brazil.}

\begin{abstract}
\noindent While standard Hamiltonian and Lagrangian methods are powerful for unconstrained classical systems, they are not directly suited to constrained ones, despite the existence of well-established treatments in the literature. In this paper, we present an alternative approach to deriving a generalized Hamiltonian and Lagrangian framework, based on the revised Barcelos-Wotzasek (BW) symplectic algorithm. This method enables a consistent and successful construction of the standard Lagrangian and Hamiltonian descriptions for constrained systems.
\end{abstract}


\maketitle
\setlength{\baselineskip}{20 pt}

\section{Introduction}

The Lagrangian formulation offers a powerful description of field theories. The deepest link between abstract mathematics and observable physics is mediated by the Lagrangian. Noether's theorem establishes an one-to-one correspondence between continuous symmetries of Lagrangian and conserved currents. This provides a rigorous field-theoretic foundation for the conservation of energy, momentum, and charge, making symmetries -- not forces -- the primary organizational principle of nature. Furthermore, interactions are not externally imposed. Through the principle of local gauge invariance, the demand that the Lagrangian remains invariant under internal symmetry groups (e.g., U(1), SU(2)) forces the replacement of ordinary derivatives with covariant derivatives. This mathematical necessity spontaneously generates interaction vertices and determines coupling constants, fully specifying how particles communicate.
Unlike the Hamiltonian, the Lagrangian is manifestly Lorentz invariant, preserving the spacetime continuum's integrity -- a prerequisite for compatibility with special and general relativity. Furthermore, it serves as the cornerstone of the path integral formulation.

This advantage naturally motivates a canonical quantization route: one begins with a Lagrangian and then passes to the Hamiltonian picture. However, this transition becomes nontrivial when the Lagrangian is singular -- in such cases, the canonical momenta cannot be inverted to solve for all velocities, giving rise to constraints within the system.

The seminal treatment of constrained systems is Dirac's constraint theory \cite{Dirac}. Subsequently, alternative phase space formalisms were developed, including those of Faddeev and Jackiw (FJ) \cite{FJ}, Barcelos and Wotzasek (BW) \cite{BW1,BW2}, and the project method \cite{CMA,CP,PM,PP}.

Our present work builds upon a revised BW formalism \cite{Ourwork}. This approach introduces a key innovation: instead of the traditional Lagrange multipliers used in the original BW scheme, it employs velocity‑-like multipliers to incorporate constraints directly at the Lagrangian level. Within the revised BW symplectic framework, the equations of motion for these velocity‑-like multipliers are computed and then substituted back into the final first‑-order Lagrangian. As a result, the extended phase space that includes the velocity‑-like multipliers reduces to the original one; the physical information originally carried by those multipliers becomes embedded in both the Lagrangian and the corresponding Hamiltonian. Consequently, the standard Euler--Lagrange equations and Hamilton's equations of motion (using Poisson brackets) can be consistently derived. At this stage, the true physical motion is obtained since  the equations of motion of the system are projected onto the constraint surface by enforcing the constraints to vanish. This procedure circumvents the operator‑-ordering ambiguities problem that typically arise in constrained Dirac quantization scheme, since the standard quantum prescription is applied:  
\[
\{A,B\} \rightarrow \frac{1}{i\hbar}[\hat A,\hat B], \quad \text{with } \Omega|\Psi\rangle = 0|\Psi\rangle,
\]  
where \(\Omega\) denotes a constraint operator.

The paper is organized as follows. Section \ref{sec2} discusses the role of the Hamiltonian within the symplectic formalism. Section \ref{sec3} formally presents the general framework for handling constrained systems. Section \ref{sec4} illustrates the methodology through some examples. Finally, Section \ref{sec5} offers conclusions and final remarks.

\section{Importance of the Hamiltonian in symplectic formalism}
\label{sec2}
The canonical Hamiltonian of a physical system is derived via the Legendre transformation of the Lagrangian; while it often corresponds to the total energy, this equivalence is not guaranteed, particularly when constraints are present. When the initial Lagrangian is singular (i.e., its Hessian matrix is degenerate), the Legendre transformation fails to be invertible, giving rise to primary constraints. To ensure physical consistency, these primary constraints must be preserved in time, which generically leads to secondary constraints. This systematic constraint analysis does not yield a ``regular Lagrangian''; rather, it enlarges the canonical Hamiltonian into the total Hamiltonian by adding Lagrange multipliers multiplied by the primary and secondary constraints, providing the Dirac framework for describing the system's constrained dynamics.

On the physical constraint surface, the system's variables lose their mutual independence. The analysis must therefore be carried out in the full, unconstrained phase space, where all canonical variables are independent. In this off-shell space, the constraints are defined as non-vanishing functions of the phase space variables; they are subsequently enforced to vanish only when projecting the final solutions onto the constraint surface.

In the symplectic formalism, the velocities (time derivatives of the fundamental coordinates) are intrinsically linked to the system's energy via the standard Hamilton's equations. Therefore, these velocities  encode primary dynamical information.

A central question arises: how can the constraint terms in the full Hamiltonian also contain information from the system's energy? The solution involves two key steps:
\begin{itemize}
\item \textbf{Work in the unconstrained framework:} This promotes the constraints from fixed conditions to dynamic, time-evolving variables.

\item \textbf{Extend the symplectic formalism:} New velocities are introduced, associated with these dynamic constraints. To achieve this, the fundamental phase space is extended by adding new coordinates whose time derivatives are these new velocities. The dynamic constraints are then elevated to the status of momenta conjugate to these new coordinates. Consequently, the new terms in the full Hamiltonian become the product of these momentum constraints and their associated velocity multipliers, analogous to kinetic terms in a first--order Lagrangian.
\end{itemize}

Given suitable conditions, one can derive Hamilton's equations for the extended phase space. These equations simultaneously determine the original velocities and the new velocity multipliers, linking their temporal evolution in a manner intimately connected to the system's energy.

\subsection{Dimensional Reduction and the Pullback of Physical Information}

The transition from the extended phase space back to the fundamental phase space is achieved through the solutions of the extended Hamiltonian system. These solutions indicate that, despite being decoupled from the fundamental variables, the velocity multipliers continue to carry nontrivial physical information derived from the energy of the system.

The physical information carried by the velocity multipliers, now recast in terms of the fundamental phase space coordinates, must be mapped into the physical system. This is achieved by substituting the explicit solutions for the multipliers and their conjugate momentum constraints directly into the full Hamiltonian. The resulting expression is a reduced Hamiltonian that depends solely on the fundamental phase space variables; it implicitly encodes the dynamical information originally contained in the multipliers, without retaining any explicit constraint terms.

This reduced Hamiltonian constitutes a new generating function in the original phase space, from which the equations of motion for the constrained system can be derived using standard unconstrained Hamiltonian methods. The final step consists of enforcing the constraints to vanish -- projecting the solutions onto the constraint surface -- so as to isolate the genuine physical equations of motion and thereby complete the dimensional reduction.

\section{General formalism}
\label{sec3}
Consider a general $n$-dimensional unconstrained system whose dynamics
is described by a Lagrangian 
\be
\label{0000}
{\cal{L}}\equiv{\cal{L}}(x_i,\dot x_i),
\ee
in which $i\in\{1,2,\dots,n\}$. The Lagrangian depends
not only on the generalized coordinates $x_i$, but also on the
generalized velocities $\dot x_i$; hence, $\cal L$ is a function on a
manifold larger than the configuration space $\mathbb {Q}$, the
\emph{velocity phase manifold} (which is not a vector space, but
the \emph{tangent bundle} or \emph{tangent manifold} \textbf{T}$\mathbb
{Q}$ of the configuration manifold $\mathbb {Q}$). 

Let us to explain this point in more detail. The \emph{tangent bundle}
\textbf{T}$\mathbb {Q}$ is obtained from $\mathbb {Q}$ by adjoining to
each point, $x_i\in\mathbb {Q}$, the tangent space
\textbf{T}$_{x_i}\mathbb {Q}$, which includes all possible velocities at
$x_i$, all tangent to $\mathbb {Q}$ at $x_i$.

The generalized momentum
$p_i$, canonically conjugated to $x_i$, is
\be 
\label{000195}
p_i=\frac{\partial {\cal{L}}(x_i,\dot x_i)}{\partial \dot x_i}.
\ee
It is possible, then, to rewrite the Lagrangian, given in Eq.(\ref{0000}), as a velocity first--order function, read as
\be 
\label{000200a}
{\cal{L}}^{(0)}=\sum_{i=1}^n p_i\dot x_i - V(x_i,p_i), 
\ee
in which $V(x_i,p_i)$ is called the Hamiltonian and the velocities $\dot x_i$ could be written as functions of the momenta $p_i$, e.g., by the
inverse of Eq.(\ref{000195}). In the FJ
symplectic algorithm, the Lagrangian ${\cal{L}}(x_i,\dot x_i)$ is
rewritten as a first--order form, given by
\be 
\label{000200b}
{\cal{L}}^{(0)}= \sum_{i=1}^{2n} A_\alpha^{(0)}\dot \xi_\alpha^{(0)} -V(\xi_\alpha^{(0)}).
\ee
in which the \textit{symplectic potential}$(V(\xi_\alpha^{(0)}))$\footnote{This term embraces the Hamiltonian at the \emph{phase manifold} \textbf{T}$^\ast\mathbb {Q}$.}, the symplectic variable and the generalized momentum
(usually called the one-form momentum) are, respectively,  $\xi_\alpha^{(0)}=(x_i,p_i)$ and $A_\alpha^{(0)}=(p_i, 0_i)$ and they belong to the
\emph{cotangent bundle} or the \emph{phase manifold} \textbf{T}$^\ast\mathbb {Q}$, which is naturally equipped with a canonical symplectic form that reproduces the Poisson bracket on ${\displaystyle \mathbb {R} ^{2n}}$. 
The Euler--Lagrange equations of motion are given by
\begin{equation}
\dot \xi_\beta^{(0)} = \left(f_{\xi_\alpha \xi_\beta}^{(0)}\right)^{-1} \partial_{\xi_\alpha^{(0)}} V(\xi_\alpha^{(0)}),
\label{EL-equation}
\end{equation}
where \(\partial_{\xi_\alpha^{(0)}} \equiv \frac{\partial}{\partial \xi_\alpha^{(0)}}\) and \(\left(f_{\xi_\alpha \xi_\beta}^{(0)}\right)^{-1}\) denotes the inverse of the symplectic matrix, which reproduces the Poisson bracket on \(\mathbb{R}^{2n}\). This symplectic matrix is defined as
\begin{equation}
\label{F000}
f_{\xi_\alpha \xi_\beta}^{(0)} = \partial_{\xi_\alpha^{(0)}} A_\beta^{(0)} - \partial_{\xi_\beta^{(0)}} A_\alpha^{(0)}.
\end{equation}
Provided that \(\det f^{(0)} \neq 0\), the matrix \(f^{(0)}\) is invertible, and \(V(\xi_\alpha^{(0)})\) is identified as the Hamiltonian. This identification reflects the fact that a nondegenerate symplectic structure ensures the velocities can be uniquely solved in terms of the canonical variables, thereby recovering the standard Hamiltonian formulation. Consequently, the pullback of the
Legendre transformation is obtained. Furthermore, Hamilton's equation of motion can be computed as well, which are
\ba 
\dot\xi^{(0)}_\alpha&=&\lbrace \xi^{(0)}_\alpha,
V(\xi^{(0)}_\alpha)\rbrace,\nonumber\\
&=&(f_{\xi_\alpha\xi_\beta}^{(0)})^{-1}\partial_{\xi_\alpha^{(0)}}V(\xi_\alpha^{(0)}),
\ea
then $\mathbb {Q}\longleftrightarrow\textbf{T}\mathbb
{Q}\longleftrightarrow\textbf{T}^\ast\mathbb {Q}$.

At this point, let us to consider a general $n$-dimensional constrained system with $m$ initial constraints arising
from the Lagrange framework or constraints of geometric nature. In the modified BW
symplectic algorithm\cite{Ourwork}, the Lagrangian ${\cal{L}}(x_i,\dot x_i)$ is
rewritten in a first--order form and the $m$ constraints are
introduced into the kinetic sector, as given by
\be 
\label{000200}
{\cal{L}}^{(0)}= \sum_{i=1}^n p_i\dot x_i +\sum_{a=1}^m \Omega_a(p_i,x_i)\dot\eta_a -V(x_i,p_i),
\ee
in which $a\in \{1,2,\dots,m\}$, $\Omega_a(p_i,x_i)$ is the $a$-th constraint and
$\dot\eta_a$ is the respective $a$-th velocity multiplier. Note that the
velocity $\dot x_i$ was {\it not} written in terms of $p_i$ and, indeed,
Eq.(\ref{000195}) might be a constraint. The first--order Lagrangian, Eq.(\ref{000200}), is rewritten as
\be 
\label{000210}
{\cal{L}}^{(0)}=\sum_{\alpha=1}^{2n+m} A_\alpha^{(0)}\dot\xi_\alpha^{(0)}  -V(\xi_\alpha^{(0)}), 
\ee
in which the symplectic variable and the generalized momentum are, respectively, $\xi_\alpha^{(0)}=(x_i,p_i,\eta_a)$ and $A_\alpha^{(0)}=(p_i, 0_i,\Omega_a)$, which belongs to the expanded
\emph{cotangent bundle} \textbf{T}$^\ast\mathbb {Q}$. We would like to emphasize that  the  \textbf{T}$^\ast\mathbb {Q}$ is expanded until its dimension becomes even. Of course, this implies that the configuration space $\mathbb {Q}$ is proportionally expanded in order to give room to the constraints inserted to the Lagrangian through velocity multipliers. After that, the symplectic matrix, given in Eq.(\ref{F000}), is computed. If $2n+m$ is odd, then $f^{(0)}$ is singular, indicating
the existence of a new constraint or a symmetry. On the other hand,
if $(2n+m)$ is even, $f^{(0)}$ can be either still singular or nonsingular.
In the latter case and in accordance with symplectic formalism \cite{BW1,BW2,Ourwork}, there will be no more constraints and the generalized
brackets among the symplectic variables are obtained from the elements of
the inverse of $f^{(0)}$. On the other hand, if the symplectic matrix
$f_{\xi_\alpha\xi_\beta}^{(0)}$ is singular, it has a zero-mode which must be contracted with the gradient of the symplectic potential. From this result, two distinct possibilities arise. The first one: this result is identically null or a previous constraint -- or combination of them -- are obtained. For this case, the system present a symmetry which is eliminated by the introduction of gauge fixing conditions (new constraint!) -- $(\Gamma_c)$ -- into the last first--order Lagrangian through a velocity multiplier $\dot\lambda_c$. The second-one: the contraction of the the zero-mode with the gradient of the symplectic potential is not null, so a new constraint $\Sigma_1$ is obtained. This constraint must be inserted
into the kinetic sector of the first--order Lagrangian, Eq. (\ref{000200}),
through a velocity multiplier $\dot\zeta_1$,
\ba  
\label{000230}
{\cal{L}}^{(1)}&=&\sum_{i=1}^n p_i\dot x_i +\sum_{a=1}^m \Omega_a\dot\eta_a +\Sigma_1\dot\zeta_1-V(\xi_\alpha^{(1)}),\nonumber\\
&=&\sum_{\alpha=1}^{M} A_\alpha^{(1)}\dot\xi_\alpha^{(1)}  -V(\xi_\alpha^{(1)}).
\ea
The new symplectic variables are
$\xi_\alpha^{(1)}=(x_i,p_i,\eta_a,\zeta_1)\equiv(x_i,p_i,\rho_\ca)$, with $\rho_\ca=(\eta_a,\zeta_1)$, the one-form momenta are $A_\alpha^{(1)}=(p_i,0_i,\Omega_a,\Sigma_1)\equiv(p_i,0_i,\Psi_\ca)$, with $\Psi_\ca=(\Omega_a,\Sigma_1)$, $\ca=1,\dots,m+1$,  $M=2n+m+1$ is even, and the new symplectic
matrix 
$f_{\xi_\alpha\xi_\beta}^{(1)}$
is computed. 
If $f_{\xi_\alpha\xi_\beta}^{(1)}$ is singular, then there will
be a new zero-mode which generates a new constraint $\Sigma_2$ and so
on. After $k$ interactions, two distinct possible scenarios may emerge. 
\begin{enumerate}
\item The
symplectic matrix $f_{\alpha\beta}^{(k)}$ is nonsingular, then the iterative procedure stops. Therefore, the
generalized brackets among the symplectic variables $\xi_\alpha^{(k)}$
are obtained from the inverse of the symplectic matrix $f_{\alpha\beta}^{(k)}$ and the
first--order Lagrangian reads as
\be 
\label{000250}
{\cal{L}}^{(k)}=\sum_{\alpha=1}^{M} A_\alpha^{(k)}\dot\xi_\alpha^{(k)}  -V(\xi_\alpha^{(k)} ),
\ee
in which $A_\alpha^{(k)}=(p_i,0_i,\Omega_a,\Sigma_b)\equiv(p_i,0_i,\Psi_\ca)$, $\Psi_\ca=(\Omega_a,\Sigma_b)$ with $b=1,\dots,k$, $\xi_\alpha^{(k)}=(x_i,p_i,\eta_a,\zeta_b)\equiv(x_i,p_i,\rho_\ca)$, $\rho_\ca=(\eta_a,\zeta_b)$ with $\ca=1,\dots,m+k$, and $M=2n+m+k$ is even.
\item The symplectic matrix
	$f_{\alpha\beta}^{(k)}$ is still singular 
	and its respective zero-mode does not generate a new constraint (or generates a previous one).
	The system has a symmetry and the zero-mode is its infinitesimal generator.
	This symmetry could be fixed as in the BW formalism. New constraints $(\Gamma_c)$ are introduced into the last first--order Lagrangian, which now is given by
\ba 
\label{000250a}
{\cal{L}}^{(k)}&=&\sum_{\alpha=1}^{M} A_\alpha^{(k)}\dot\xi_\alpha^{(k)}-V(\xi_\alpha^{(k)} ),\nonumber\\
&=&\sum_{\alpha=1}^{2n} A_\alpha^{(0)}\dot\xi_\alpha^{(0)}+ \underbrace{\sum_{\ca=1}^{M}\Psi_\ca\dot{\rho}_\ca-V(\xi_\alpha^{(k)})}_{-\cal{H}} ,
\ea
in which $A_\alpha^{(k)}=(p_i,0_i,\Omega_a,\Sigma_b,\Gamma_c)\equiv(p_i,0_i,\Psi_\ca)$, $\Psi_\ca=(\Omega_a,\Sigma_b,\Gamma_c)$ with $c=1,\dots,s$, $\xi_\alpha^{(k)}=(x_i,p_i,\eta_a,\zeta_b,\lambda_c)\equiv(x_i,p_i,\rho_\ca)$, $\rho_\ca=(\eta_a,\zeta_b,\lambda_c)$ with $\ca=1,\dots,m+k+s$, and $M=2n+m+k+s$ is even.	
\end{enumerate}
Afterthat, the Euler-Lagrange equation of motion is computed
\be
\label{Vec_EqMot}
\dot \xi_\beta^{(k)}=(f_{\xi_\alpha^{(k)}\xi_\beta^{(k)}})^{-1}\partial_{\xi_\alpha^{(k)}}V(\xi_\alpha^{(k)})
\ee

Although the full Hamiltonian is, effectively, given by ${\cal{H}}={V-{\Psi}\cdot{\dot\rho}}$, as can be seen in Eq.(\ref{000250a}), the methodology processes the energy of the physical system $\,V\,$ and the constraint term $\,-\Psi\cdot\dot\rho\,$ differently, in such a way that the energy of the physical system in Eq.(\ref{Vec_EqMot}) remains in $\,{\d{}V}/{\d\xi}\,$ while the constraint term is transferred to the Dirac bracket matrix $\,f^{-1}\,$ as part of the kinetic term. 
Then, the components of Eq. (\ref{Vec_EqMot}) are the separate equations of motion for all coordinates of the extended phase space $\,(\q, \p, \rho)$. 
So, the latter $\,M\,$ components of the vector $\,\dot\xi\,$ in Eq. (\ref{Vec_EqMot}) express the adjustment of the velocity multipliers $\,\dot\rho\,$ to the physical system through the energy of the system, then they will give us the velocity multipliers $\,\dot{\rho}(x^i,p^i)$, loaded with the information about the physical system, as a function of $\,x^i$, $p^i$. 
To find the reduced Hamiltonian -- a Hamiltonian written in terms of phase space coordinates $\,(\q, \p)$ -- the velocity multipliers $\dot{\rho}(x^i,p^i)$, obtained from the Euler-Lagrange equations, Eq. (\ref{Vec_EqMot}), must be substituted into the full Hamiltonian.
Therefore, the reduced Lagrangian could be computed from Eq.(\ref{000250a}) and expressed as a Legendre transformation,
\be
{\cal L}=\p\cdot\dot{\q}-{\cal H}.  \label{LagOutDef}
\ee

With the dimensional reduction of the phase space already completed, we compute $\,\dot\q\,$ and $\,\dot\p\,$ using the canonical Hamilton's procedure in order to get the equations of motion. From these equations and Eq.(\ref{LagOutDef}), the reduced Lagrangian $({\cal L})$ is expressed only in terms of $x^i$ and $\dot{x}^i$, which is our goal. In this way we were drove to a system in a reduced phase space in which the $\,2n\times2n\,$ matrix $\,f^{-1}\,$ is the standard symplectic matrix whose elements are the Poisson brackets that, through the canonical quantization rule, lead us directly to the canonical commutation relations of Quantum Mechanics, so the degrees of freedom are the physical ones. It is noteworthy that the constraints $\,\Psi_\ca\,$ were not taken equal to zero at any time. Thus, the Lagrangian and Hamiltonian that was obtained refers to a state prior to that projected onto the constraint surface. So, both the Euler-Lagrange equation applied to the Lagrangian  and the Hamilton's equations applied to the Hamiltonian resuls in the same equation of motion of the unconstrained framework. Then, in  order to obtain the true motion, the equation of motion of the physical system must be projected onto the constraint surface by forcing the constraints to take zero values.

\section{Applications}
\label{sec4}
In order to illustrate the methodology presented in Section \ref{sec3}, this method is applied in two constrained systems: the hypersphere and the free electromagnetic field.

\subsection{Hypersphere }
A nonrelativistic particle of mass $\,m\,$ moving on the surface of an $n$-dimensional sphere (hypersphere) of constant radius $\,R\,$ has an intrinsic constraint, given by
\be 
\Omega_1=\frac12\left(\q^2-R^2\right),
\ee 
regarding geometry, and its dynamics is governed by the following Lagrangian
\be 
{\cal L}=\frac {m\dot x_i^2}{2},
\ee
with the conjugate canonical momentum $\,\p=m\dot \q$ to $\q$. Applying the symplectic BW revised, well demonstrated in subsection III.A of \cite{Ourwork}, the last first-order Lagrangian is obtained as
\be
{\cal L}=\p\cdot\dot\q\overbrace{+\Omega_1\,\dot\eta_1+\Omega_2\,\dot\eta_2-U}^{-{\cal H}},   \label{LagHySp}
\ee
with
\be
U=\frac{\p^2}{2m}\,, ~~~~   ~~~~\text{and}~~~~ \Omega_2=\frac{\q\cdot\p}m,     \label{hypersphere}
\ee
where $U$ is the Hamiltonian and, $\,\Omega_1$ and ${\Omega}_2$, are constraints. 

Expressing Eq.(\ref{LagHySp}) in a symplectic form, we readily identify the $(2n+2)\times1$ symplectic vector $\,\xi\,$ of the extended phase space and its conjugate momentum vector $\,a\,$ as
\be
\arraycolsep=1.4pt\def\arraystretch{1.3}
\xi=\left(
\begin{array}{c}
\q     \cr
\p     \cr
\eta_1 \cr
\eta_2  
\end{array}\right),
\qquad
\arraycolsep=1.4pt\def\arraystretch{1.3}
a=\left(
\begin{array}{c}
\p          \cr
\o           \cr
\frac12\left(\q^2-R^2\right)    \cr
\frac{\q\cdot\p}m
\end{array}\right).   \label{PS_FJHyp}
\ee
Then, using the matrix structure of Eq.(\ref{F000}), we get
\be
\arraycolsep=1.4pt\def\arraystretch{1.3}
{f}=\left(
\begin{array}{cccc}
\O            ~&~     -\I       ~&~ \q ~&~ ~\frac{\p}m  \cr
\I            ~&~     ~\O       ~&~ \o ~&~ ~\frac{\q}m  \cr
-\q^T         ~&~   ~~\o^T      ~&~  0 ~&~     ~0       \cr
-\frac{\p^T}m ~&~ -\frac{\q^T}m ~&~  0 ~&~     ~0
\end{array}\right).  \label{FJ1-matrix}
\ee
The inverse of $\,{f}\,$ is 
\be
\arraycolsep=1.4pt\def\arraystretch{1.3}
{f}^{-1}=\left(
\begin{array}{cccc}
\O                        ~&~ \I-\frac{\q\,\q^T}{\q^2}       ~&~ -\frac{\q}{\q^2} ~&~ \o                \cr
-\I+\frac{\q\,\q^T}{\q^2} ~&~ \frac{\p\,\q^T-\q\,\p^T}{\q^2} ~&~ \frac{\p}{\q^2}  ~&~ -\frac{m\q}{\q^2} \cr
\frac{\q^T}{\q^2}         ~&~ -\frac{\p^T}{\q^2}             ~&~ 0                ~&~ -\frac{m}{\q^2}   \cr
\o^T                      ~&~ \frac{m\q^T}{\q^2}             ~&~ \frac{m}{\q^2}   ~&~ 0
\end{array}\right).  \label{DBmatrix}
\ee
Note that $\,\q^T\q = \q^2\,$ is a scalar and $\,\q\,\q^T$, $\p\,\q^T$, $\q\,\p^T\,$ are $n\times{}n$ matrices. 
So, the Dirac brackets of the coordinates of the fundamental phase space can be read from the elements of the matrix, Eq.(\ref{DBmatrix}), as
\ba
\label{DBqq}
\left\{x_i~,~x_j\right\}_{\!_D}&=&0,                              
\nonumber\\
\left\{x_i~,~p_j\right\}_{\!_D}&=&\delta_{ij}-\frac{x_i\,x_j}{\q^2}, \\
\left\{p_i~,~p_j\right\}_{\!_D}&=&\frac{p_i\,x_j-x_i\,p_j}{\q^2}.  \nonumber 
\ea
In the same way, we can also easily read the Dirac brackets involving the extra coordinates associated with the constraints. At the quantum level, these brackets, Eq.(\ref{DBqq}), present the  operator‑-ordering ambiguities problem at the quantum level.

Next, the matrix expression for the velocities of the extended phase space, given in Eq.(\ref{Vec_EqMot}), with the Dirac bracket matrix, Eq.(\ref{DBmatrix}), and the following vector
\be
\arraycolsep=1.4pt\def\arraystretch{1.3}
\frac{\d{}U}{\d\xi}=\left(
\begin{array}{c}
\o   \cr
\p/m \cr
0    \cr
0
\end{array}\right),  \label{DH0_sph}
\ee
provides the velocity multipliers
\be
\dot{\eta_1} = -\frac{\p^2}{m\q^2} \quad\text{and}\quad \dot{\eta_2} = \frac{\q\cdot\p}{\q^2}.        \label{HamEq_eta1,2}                                  
\ee
Thus, by using Eq.(\ref{hypersphere}) and Eq.(\ref{HamEq_eta1,2}), the Hamiltonian $({\cal H})$, Eq.(\ref{LagHySp}), changes to
\be
{\cal H}=\frac{\p^2}{2m}\left(2-\frac{R^2}{\q^2}\right)-\frac{\left(\q\cdot\p\right)^2}{m\q^2},    \label{HamOut}
\ee
which is the reduced Hamiltonian. At this point, the constraints's contribution is embedded into the Hamiltonian. Furthermore, the first-order Lagrangian, Eq.(\ref{LagHySp}), through symplectic formalism, generates the Poisson brackets among the phase space variables, namely:
\be 
\label{POBB}
\lbrace \q,~\q\rbrace=\lbrace \p,~\p\rbrace=0, ~~\lbrace \q,~\p\rbrace=\I.
\ee
In this scenario, these brackets, Eq.(\ref{POBB}), do not present operator‑-ordering ambiguities problem at the quantum level.

From  the reduced Hamiltonian, Eq.(\ref{HamOut}), the equations of motion, the new canonical momentum and the reduced Lagrangian expressed in terms of $\q$ and $\dot\q$ can be calculated. The canonical Hamilton's equations of motion are
\ba
\dot\q &=& \lbrace\q,~{\cal H}\rbrace=\frac{\p}m\left(2-\frac{R^2}{\q^2}\right)-\frac{2\left(\q\cdot\p\right)}{m\q^2}\,\q,   \label{firstHamEq}
\\&&\nn\\
\dot\p &=&\lbrace\p,~{\cal H}\rbrace= -\frac{\p^2R^2}{m\left(\q^2\right)^2}\,\q+\frac{2\left(\q\cdot\p\right)}{m\q^2}\left[\p-\frac{\left(\q\cdot\p\right)}{\q^2}\,\q\right].       \label{secondHamEq}
\ea
The canonical momentum is obtained from Eq.(\ref{firstHamEq}) as being
\be
\p=\left[m\dot\q+\frac{2\left(\q\cdot\p\right)}{\q^2}\,\q\right]\left(\frac{\q^2}{2\q^2-R^2}\right).  \label{mom_from_Ham}
\ee
One relation that helps us to map the phase space to configuration space is obtained with the scalar product of  Eq.(\ref{mom_from_Ham}) by $\q$, namely:
\be
\q\cdot\p=-\left(\frac{m\q^2}{R^2}\right)\q\cdot\dot\q.     \label{qcdotp}
\ee
By a straightforwad inspection of Eq.(\ref{HamOut}) and Eq. (\ref{firstHamEq}), it is possible to express the Hamiltonian, Eq. (\ref{HamOut}), as
\be
{\cal H}=\frac12\,\p\cdot\dot{\q}.  \label{1orderHam}
\ee
From the Legendre transformation, Eq.(\ref{LagOutDef}), and using the Hamiltonian, Eq.(\ref{1orderHam}), the first--order Lagrangian is obtained as being
\be
{\cal L}=\frac12\,\p\cdot\dot{\q}.   \label{1orderLag}
\ee
From Eq.(\ref{mom_from_Ham}) and Eq.(\ref{qcdotp}), the Lagrangian of the unconstrained framework, given in Eq.(\ref{1orderLag}), is expressed in terms of the coordinate $\,\q\,$ and velocity $\,\dot\q\,$, which is read as
\be
{\cal L}=\frac12\,m\left[\dot\q^2-\frac{2\left(\q\cdot\dot\q\right)^2}{R^2}\right]\left(\frac{\q^2}{2\q^2-R^2}\right), \label{LagOut}
\ee
whose the determinant of Hessian matrix is $det H=-m^n$, which means a  nonconstraint model. Consequently, the Legendre transformation is restored.

The Euler-Lagrange equation applied to the Lagrangian, given in Eq.(\ref{LagOut}), results in the equation of motion of the unconstrained framework,
\be
\ddot\q=-\frac{{\dot\q}^2}{\q^2}\,\q-\frac{2R^2\left(\q^2-R^2\right)}{\q^2\left(2\q^2-R^2\right)^2}\left[{\dot\q}^2-\frac{\left(\q\cdot\dot\q\right)^2}{\q^2}\right]\q\,, \label{EqOfMotOut}
\ee
which is the same one obtained by the Hamilton's equations of motion, given in Eq.(\ref{secondHamEq}).
Then, the equation of motion, Eq.(\ref{EqOfMotOut}), must be projected onto the constraint surface by imposing the constant radius constraint, which is expressed by the constraint $\Omega_1$, so the equation of motion changes to
\be 
\ddot\q=\frac{\dot{\q}^2}{R}\left(-\frac{\q}R\right).   \label{EqOfMotOn}
\ee
In this last, the unit vector $\,\q/R\,$ is perpendicular to the trajectory of constant radius $\,R$, as can be verified by Eq. (\ref{qcdotp}) taking the orthogonality constraint equal to zero, and the expression $\,\dot{\q}^2/{R}\,$ is the magnitude of the centripetal acceleration $\,\ddot\q\,$ of the particle of mass $\,m$, which  points towards the center of the hypersphere of radius $R$. Therefore, the particle of mass $\,m\,$ can follow paths on any great hypercircumference on the hypersphere surface of radius $\,R$. 

\subsection{Free Electromagnetic Field}

In this Section, we consider the  metric $\,\eta_{\mu\mu}=(+1~-1~-1~-1)$. As $\,x^0=t\,$ is the time and $\,x^i$ are the coordinates of the position $\,\x$, the spacetime coordinates and derivatives are given by
\ba
x^{\mu}=(t~,~\x)                     \quad&,&\quad	x_{\mu}=(t~,~-\x) ~,\\
\d_{\mu}=(\frac{\d}{\d t}~,~\nabla)  \quad&,&\quad	\d^{\mu}=(\frac{\d}{\d t}~,~-\nabla) ~.
\ea
We denote the \textit{partial derivative} with respect to time of a quantity by a dot above it. From the electromagnetic tensor $\,F^{\mu\nu}=\d^\mu{}A^\nu-\d^\nu{}A^\mu$, we gather the electric field, ${\E=-\dot{\A}-\nabla{}A^0\,}$ and the magnetic field, $\B=\nabla\times\A$. On the other hand, from the Lagrangian density
\be
\mathcal{L}=-\frac14F_{\mu\nu}F^{\mu\nu}.    \label{lag_emwaves}
\ee
From the implementation of symplectic BW formalism revised in Eq.(\ref{lag_emwaves}), the last first-order Lagrangian density is obtained as being
\be
\mathcal{L}=-\pib\cdot\dot\A+\Omega_1\dot\eta_1-\mathcal{U},   \label{lagconst1}
\ee
where a new constraint is obtained, the Gauss's law $(\Omega_1=\nabla\cdot\pib)$, and is identified an existence of a gauge symmetry. The energy density of the electromagnetic waves, expressed in terms of the coordinates of the fundamental phase space $\,(\A, \pib)\,$, as
\be
\mathcal{U}=\frac12\,\pib^2+\frac12\left(\nabla\times\A\right)^2.    \label{emenergy} 
\ee
To fix the symmetry, a gauge fixing condition $(\Omega_2=\nabla\cdot\A)$ is introduced into Eq.(\ref{lagconst1}), which is now given by
\be
\mathcal{L}=-\pib\cdot\dot\A\overbrace{+\Omega_1\dot\eta_1+\Omega_2\dot\eta_2-\mathcal{U}}^{-{\cal H}},   \label{lagconst}
\ee
The symplectic vector $\,\xi\,$ of the extended phase space  and its conjugate momentum vector $\,a\,$ are given by
\be
\arraycolsep=1.4pt\def\arraystretch{1.3}
\xi=\left(
\begin{array}{c}
\A \cr
\pib   \cr
\eta_1 \cr
\eta_2  
\end{array}\right),
\qquad
\arraycolsep=1.4pt\def\arraystretch{1.3}
a=\left(
\begin{array}{c}
-\pib           \cr
\o              \cr
\Omega_1 \cr
\Omega_2
\end{array}\right).   \label{PS_FJem}
\ee
In order to compute the equations of motion, Eq.(\ref{Vec_EqMot}), the symplectic matrix and its inverse are computed, namely:
\be
\arraycolsep=1.4pt\def\arraystretch{1.3}
{f}=\left(
\begin{array}{cccc}
\O       ~&~   \I       ~&~   \o    ~&~ -\nabla \cr
-\I      ~&~   \O       ~&~ -\nabla ~&~      \o \cr
\o^T     ~&~  \nabla^T  ~&~    0    ~&~       0 \cr
\nabla^T ~&~   \o^T     ~&~    0    ~&~       0
\end{array}\right),  \label{fmatrix}
\ee
whose its inverse is
\be
\arraycolsep=1.4pt\def\arraystretch{1.3}
{f}^{-1}=\left(
\begin{array}{cccc}
\O                                   ~&~ -\I+\frac{\nabla\,\nabla^T}{\nabla^2} ~&~ \o                       ~&~ \frac{\nabla}{\nabla^2} \cr
\I-\frac{\nabla\,\nabla^T}{\nabla^2} ~&~       \O                              ~&~ \frac{\nabla}{\nabla^2}  ~&~                     \o  \cr
\o^T                                 ~&~   -\frac{\nabla^T}{\nabla^2}          ~&~ 0                        ~&~       -\frac1{\nabla^2} \cr
-\frac{\nabla^T}{\nabla^2}           ~&~      \o^T                             ~&~ \frac1{\nabla^2}         ~&~                      0
\end{array}\right).  \label{DBmatrixem}
\ee
The velocity vector of the extended phase space is,
\be
\dot\xi(\x)=\iiint{}d^{3}\y\,f^{-1}(\x-\y)\,\frac{\d\,\mathcal{U}}{\d\xi(\y)}.
\ee
However, the Dirac bracket matrix is
\be
f^{-1}(\x-\y)=\delta^{(3)}(\x-\y)\,f^{-1}(\y).
\ee
Then, we can write the matrix equation for the velocity vector as a local expression,
\be
\dot\xi=f^{-1}\,\frac{\d\mathcal{U}}{\d\xi}.     \label{localvel}
\ee
From the energy density, given in Eq.(\ref{emenergy}), we get
\be
\arraycolsep=1.4pt\def\arraystretch{1.3}
\frac{\d\mathcal{U}}{\d\xi}=\left(
\begin{array}{c}
\o   \cr
\pib \cr
0    \cr
0
\end{array}\right).  \label{DH0}
\ee
From the product of the matrix $\,f^{-1}\,$, Eq.(\ref{DBmatrixem}),  with the vector, Eq.(\ref{DH0}), we get, as indicated by Eq.(\ref{localvel}), the velocity multipliers
\ba
\dot\eta_1&=&-\frac{\nabla\cdot\pib}{\nabla^2} \quad\text{and}\quad    \label{velmulem}
\dot\eta_2=0.
\ea
The Lagrangian density, given in Eq.(\ref{lagconst}), can be written as the Legendre transformation,
\be
\mathcal{L}=-\pib\cdot\dot\A-\mathcal{H},  \label{fieldLegendre}
\ee
where $\,\mathcal{H}\,$ is
\be
\label{hhh}
\mathcal{H}=\mathcal{U}-\left(\nabla\cdot\pib\right)\dot\eta_1-\left(\nabla\cdot\A\right)\dot\eta_2.
\ee
Considering the velocity multipliers, given in Eq.(\ref{velmulem}), the full Hamiltonian density, Eq.(\ref{hhh}), of the electromagnetic waves changes to the reduced Hamiltonian in the unconstrained framework, read as
\be
{\cal H}=\frac12\,\pib^2+\frac12\left(\nabla\times\A\right)^2+\frac{\left(\nabla\cdot\pib\right)^2}{\nabla^2}.   \label{ham}
\ee
At this point, the canonical Hamilton's equations of motion are computed, namely:

\ba
\dot\A   &=& -\pib+2\,\frac{\nabla\left(\nabla\cdot\pib\right)}{\nabla^2},  \label{sdotmom2}
\\&&\nn\\
\dot\pib &=& -\left[\nabla^2\A-\nabla\left(\nabla\cdot\A\right)\right]. \label{sdotvec2}
\ea
Applying the divergence operator on Eq.(\ref{sdotmom2}), the relation that maps the phase space to the configuration space is obtained
\ba
\nabla\cdot\dot\A  &=& \nabla\cdot\pib.   \label{PhaseToConfig}
\ea
So, using Eq.(\ref{sdotmom2}) and Eq.(\ref{PhaseToConfig}), we obtain that the conjugate momentum to $\,\A$ in the unconstrained framework, given by
\be
\pib=-\dot\A+2\,\frac{\nabla\left(\nabla\cdot\dot\A\right)}{\nabla^2}.   \label{newmom}
\ee
In order to obtain the equation of motion of the electromagnetic waves, we can combine Eq.(\ref{sdotvec2}) with Eq.(\ref{newmom}) to eliminate $\,\dot\pib$, thus
\be
\ddot\A=\nabla^2\A-\nabla\left(\nabla\cdot\A\right)+2\,\frac{\nabla\left(\nabla\cdot\ddot\A\right)}{\nabla^2}.  \label{eqmotham}
\ee
Using Eq.(\ref{ham}), Eq.(\ref{PhaseToConfig}) and Eq.(\ref{newmom}), the Lagrangian density, given in Eq. (\ref{fieldLegendre}) changes to the reduced Lagrangian in the unconsrained framework, given by
\ba
{\cal L}&=&\frac12\dot\A^2-\frac12\left(\nabla\times\A\right)^2-\frac{\left(\nabla\cdot\dot\A\right)^2}{\nabla^2}-2\left[\frac{\nabla\left(\nabla\cdot\dot\A\right)}{\nabla^2}\right]^2.   \label{newlag}
\ea
The reduced Lagrangian, Eq.(\ref{newlag}), and Hamiltonian, Eq.(\ref{ham}), were then derived in a straightforward way.

The longitudinal and transverse components of vector fields $\,\A\,$ and $\,\pib$ are given by
\ba
\A_\L=\frac{\nabla\left(\nabla\cdot\A\right)}{\nabla^2}\,,&\qquad&\A_\T=-\frac{\nabla\times\left(\nabla\times\A\right)}{\nabla^2}\,,
\\&&\nn\\
\pib_\L=\frac{\nabla\left(\nabla\cdot\pib\right)}{\nabla^2}\,,&\qquad&\pib_\T=-\frac{\nabla\times\left(\nabla\times\pib\right)}{\nabla^2}\,,
\ea
with the following algebra:
\ba
\A=\A_\L+\A_\T\,;&\qquad& \pib=\pib_\L+\pib_\T;
\\&&\nn\\
\pib_\T\cdot\A_\L=0\,;&\qquad&\A_\T\cdot\pib_\L=0;
\\&&\nn\\
\A_\T\cdot\A_\L=0\,;&\qquad&\pib_\T\cdot\pib_\L=0.
\ea
Next, dropping a surface term, the Lagrangian density, Eq. (\ref{newlag}), can be expressed as
\ba
{\cal L}&=&\frac12\dot\A^2-\frac12\left(\nabla\times\A\right)^2-\dot\A_\L^2.   \label{newlag2}
\ea
The Euler-Lagrange equation for vector fields is
\be
\frac{\d}{\d{}t}\,\frac{\d {\cal L}}{\d\,\dot{\A}}+\d_j\,\frac{\d{\cal L}}{\d\,\d_j{}\A}-\frac{\d {\cal L}}{\d\,\A}=\o. \label{sE-L}
\ee
Since
\be
\frac{\d {\cal L}}{\d\,\dot{\A}}=\dot\A-2\,\dot\A_\L, \qquad \frac{\d {\cal L}}{\d\,\A}=\o.    \label{twovecderL}
\ee
Using the Euler-Lagrange equation, Eq.(\ref{sE-L}), with the Lagrangian density, Eq.(\ref{newlag2}), we get  
\ba
\ddot\A&=&\nabla^2\A-\nabla\left(\nabla\cdot\A\right)+2\,\ddot\A_\L,\nonumber
\\
 \ddot\A&=&\nabla^2\A_\T+2\,\ddot\A_\L, \label{eqmotlag}
\ea
which is the same equation of motion, given in Eq.(\ref{eqmotham}). Spliting the last equation of motion, given in Eq.(\ref{eqmotlag}), into transverse and longitudinal sectors, we obtain
\ba
\ddot\A_\T-\ddot\A_\L&=&\nabla^2\A_\T.
\ea
This is the equation of motion in the unconstrained framework. Let us project it onto the constraint surface, where the transversality constraint $\,\A_\L=\o\,$ holds, in a way that
\be
\ddot\A=\nabla^2\A.
\ee
This is the wave equation for the electromagnetic field $\,\A$, which is the expected result to confirm the formalism.

\section{Conclusion}
\label{sec5}
Conventional approaches to constrained gauge systems frequently compromise essential properties, undermining manifest Lorentz invariance and violating the strict Dirac quantization map from Poisson brackets to commutators. To overcome these inherent difficulties, we propose an alternative constructive method for deriving a Lagrangian that systematically incorporates all contributions stemming from the constraints. Crucially, this method restores the full Legendre transformation $
\mathbf{T}\mathbb{Q} \longleftrightarrow \mathbf{T}^*\mathbb{Q}$,
enabling a proper Hamiltonian formulation on the reduced phase space that is canonically associated with the original Lagrangian.

The resulting dynamics are entirely encoded in a canonical symplectic form represented by a $2n \times 2n$ matrix $(f^{-1})$, whose elements define the Poisson bracket structure on $\mathbb{R}^{2n}$. Under Dirac's canonical quantization prescription, these brackets directly generate the canonical commutation relations of quantum mechanics. Consequently, the operator-ordering ambiguities are consistently shifted to the Hamiltonian sector, the fundamental commutators remain strictly canonical, faithfully mirroring the classical Poisson structure. Furthermore, the reduced Lagrangian is obtained, then the path integral quantization procedure becomes readily implementable within our framework.

%=====================================================================
\begin{itemize} 
\item[] \textbf{Author Contributions~} All authors contributed equally to the elaboration of this manuscript, from its conception to its final form.
\item[] \textbf{Funding~} This work is partially supported by FAPERJ and CNPq (Brazilian Research Agencies).
	E.V. Corr\^{e}a Silva thanks Universidade do Estado do Rio de Janeiro (UERJ) for the Proci\^{e}ncia grant.
\item[] \textbf{Data Avaliability~} All of the material is owned by the authors and/or no permissions have been required. No data sets have been generated during the current study.
\end{itemize}
%=====================================================================
\section*{Declarations}

\begin{itemize}
\item[] \textbf{Ethical Approval~} This manuscript complies with the ethical policies of the journal.
\item[] \textbf{Competing Interests~} The authors declare no competing interests.
\end{itemize}

\end{document}